# ON THE EVOLUTION OF STRUCTURE, IN TURBULENCE PHENOMENA

Luís G.D. de Calvão Borges

**Abstract.** We study the evolution of coherent structures in arbitrary turbulence phenomena, developing some tools, from non-archimedean analysis and algebraic geometry, in order to model its display. We match the scale-dependent, topological structure of the physical fields subsumed by these coherent structures with the phase portraits of some convenient, non-archimedean dynamical systems whose set we endow with a law of action. Finally, we obtain the proposition of an extremum principle for this action to define a geodesic path, as the modelized coherent structures evolve, and we delineate the conception of two experiments, derived from the fields of fluid and plasma turbulence, to test its efectiveness.

# 1. Introduction.

We expose a survey of unpublished research on the evolution of structure in turbulence, in whatever form that it presents, or developes itself in the world but, formally, understood as the problem of studying the ensemble of solutions to an arbitrary, deliberately unknown system of differential equations.

Real world examples of turbulence come, traditionally, from fluids described by the Navier-Stokes equations [32,40,60,73,91], although one may recognize the same, problematic situation when studying the magnetic confinement of plasmas, the fault dynamics in the earth's crust, or even the behavior of markets in an economy [10,33,46,50,65,86,90,92]. Now, facing such a generic sort of phenomena, our intuition is impressed by the prevalence of scale invariance that it is, concomitantly, exhibited. And the idea follows that something, like structure, should be conserved, as such a symmetry prevails; an idea that, as it is well known, has its roots in the work of Noether [69], in 1918. But, there is no exact scale invariance in any real world phenomenon, whatsoever, and so, we are naturally led to consider the question of determining the flow of that something, like structure, that, in fact, it is not conserved.

Aiming for objectiveness, we concentrate on the emergence and the evolution of coherent structures [40,66,87,88], as they are exhibited, for example, by vortices in a turbulent flow. The scope of our analysis unfolds, therefore, for us to face both the many-scale problem with strong coupling between the constituent scales and the problem of the connection between scaling and structure [91]. We observe such a task to be a feasible one, as we manage to develop a conceptual framework where both these two problems, and the main challenge that originated them, can all be treated within the rigorous terms of a mathematical formalization that generates models encompassing their respective, possible phenomenological manifestations, whatever they may be. And some consequences follow, allowing the efectiveness of this framework to be put to the trial of experience. For such an efectiveness to obtain, a few options have to be made, at the expense of a rather elementary use of some, quite demanding mathematics, notwithstanding the fact that, in what concerns modelization methodology, we do not depart significantly from current practices.

We start with the common observation that many turbulent flows display coherent structures across several orders of spatial or temporal magnitude . We take this fact as evidence for the concomitant involution of something quite likely to be expressed, by analogy, as a renormalization flow [17], of which the observed coherent structures constitute the snapshots. Consequently, we identify these snapshots as the scale-dependent state representations of what we call Sigma structures, and we study their evolution under a certain, convenient law of action, $\Psi$, having scale as its parameter. It is fundamental to acknowledge that we must leave the (physico-mathematical) reference, or the (ontological) status of these Sigma structures undetermined [22,59].

Following these guidelines, we consider the specifics of a particular turbulent fluid flow and the associated Sigma structure. As we stated, the display of coherent structures at a given order of, say, spatial magnitude is to be interpreted as a scale-dependent state representation of the Sigma structure. Now, we postulate that such a state representation is faithfully modelized by a convergent power series, with integer coefficients, restricted to be defined on some subset of the maximal ideal of a complete, algebraically closed, local field containing the rational numbers [11,56,77,81]. The modeling procedure is acomplished in three, interwoven steps: to begin with, the topological structure of the vorticity field associated with the turbulent fluid flow must be obtained [73,79]; then, an euclidean model is given to the p-adic integer ring contained in the ground number field [5,20,76,77]; finally, an adjustment is to be sought for, between the topological structure of the physical field and the phase portrait of an adequate power series, on the given euclidean model.

Thereafter, a certain, convenient law of action, or operation on the set, $Z_p[x]$, defined by these power series is taken as the central object of our study. Starting from the postulate that it must satisfy an extremum principle, in fact, a minimum one, we will prove, first, that it is unique, if it exists and if it is subjected to the proper, formal constraints; afterwards, it will be shown that it does exist and, finally, that it factors in a very special, also unique way.

The demonstration of these three main results constitutes the fulcrum of our research. In particular, the strategy that we adopt in order to achieve them, exploring the various levels of mathematical structure that the chosen power series veil, reveals itself as a sucessful one, as it authorizes us on the far-reaching assertion that *the evolution of structure in turbulence phenomena follows some well determined paths, that are geodesic, in a precise, experimentally testable way.*

After these introductory lines, our survey is organized in the following way:

In section 2, there is a preliminary discussion, aiming to set the context and the motivation for the achievement of our first main result; then, we show that, if there exists a law of action, or operation on the set $Z_p[x]$, and if it is restricted by some proper, bottom-up, scale-dependent formal constraints, then, the result of that operation upon a given, initial power series, $u(x)$, which itself is restricted to be defined only up to a certain order of magnitude, is a unique one.

In section 3, we deal with the problem of existence for the law of action that was previously dealt with, showing, in particular, that it can be qualified as one of analytic extension through T-filters, between power series defined on distinct subsets of a totally disconnected space; this is our second main result.

In section 4, after another small, preliminary discussion, we show that such an action admits a unique, non-trivial factorization along morphisms in the category of rigid analytic varieties, exposing, thereby, the third and last of our main, mathematical results.

Sections 5 and 6 will debate the apparently innocuous character of our factorization result; their presentations are somewhat terse, as we wish not to dwell on the further mathematics called upon by the analysis of the strategy of proof that was adopted before, in section 4. Instead of that, in section 5, we offer an informal illustration of some connection between our work and, both, number theory and dynamical systems theory, and afterwords, in section 6, we elaborate on the inference of the claim that we submit to an eventual, possible test in the laboratory, that is to say, the proposition that the development of turbulences follows the proper geodesics that we define in an auxiliary space.

In section 7, this survey is driven to its end, with a few, final considerations.

For the convenience of an eventual reader not familiar with the variety of mathematical theories that are summoned along our survey, and/or the physics to which they are applied, we have chosen to be somewhat redundant on the references (section 8) that are given along the text. In what concerns notation, we have followed standards, whenever not explicitly stated otherwise.

Finally, we observe that the proofs of theorems 1 and 2 may be skipped, on a first reading of this survey, without losing grasp of the core of its physical content; on the contrary, the proof of theorem 3 is more important than the theorem itself.

## 2 - Uniqueness under the proper constraints.

Let there be given some physical system undergoing turbulence, and let it be such that it allows for the display of a certain number of scale invariant coherent structures across several orders of magnitude.

We will assume that we have the knowledge of some, definite, otherwise arbitrary models of the scale-dependent states of these coherent structures, up to a certain, fixed low order of magnitude. The model that we consider to be defined at the greatest of these orders of magnitude, and that is the model that we take as the initial state for our evolution problem, is a convergent power series, $u_n(x)$, written

$$u_n(x) = x^3 (1 + A_{n,1} x + A_{n,2} x^2 + ...) ,$$

with its coefficients in $Z_p$, the ring of the p-adic integers. The functional form of $u_n(x)$ is suposed to be valid only in the set that is the inverse image, $V_n$, defined as

$$V_n = u_n^{-1} (p^n Z_p \setminus p^{n+1} Z_p) ,$$

where n stands for a natural number, p stands for an arbitrary prime number, and $p^n Z_p$ denotes the set $\{x : x \in Z_p , |x|_p < p^{-n}\}$ [11,56,77,81,83].

We will assume, also, that we are given the formal statement of the relevant boundary conditions constraining the evolution of coherent structures, from that initial state onto some greater orders of magnitude. In particular, these bottom-up constraints are thought to embody the demand for the law of action on $Z_p[x]$ to be minimal upon $u_n(x)$ and, consequently, to ensure that there will be some, quite well determined permanence of structure along with that operation (see [49], chapts 9 and 10, and [13], Note Historique, IV.49-IV53).

We make a simple, formal logico-algebraic proposition out of these considerations that will be shown to entail uniqueness for the action upon $u_n(x)$. The demonstration that it indeed happens to be so, as we chose to deliver it, is technically dependent, first, on a simple application of the topological duality established by Stone [95-97], then, after some elaboration, on the use of some consequences derived from the pragmatic foundations of standard logic, as proposed by Sweet [98-101], and finally, on the resort to a particular property of the polyadic algebras introduced by Halmos [34-37], as exposed by Daigneault and Monk [21].

Now, let us consider the observation of an arbitrary turbulence phenomenon as the one that was suggested above. Such an observation is bound to proceed up to a certain, protocol dependent cut-off scale, and then stops, but we assume that the experimental data thereby obtained can be used, in particular, to infer a certain system of differential equations, modeling the phenomenon, the specifics of which we deliberately ignore.

This inference is meant to be made by a set of idealized users [98,99] of the formal inner language, IL, of the relevant physical theory [15], which is supposed to be completely known to them.

Subsequently, there results the knowledge, both of a sequence of mathematical objects, $(u_i)_{i=1,2,...,n}$, modeling the state of the observed coherent structures, respectively, at each scale inside the window of observation, and of a certain set of formal clauses, $\Delta$, expressing the formal boundary conditions that are constraining the ongoing development of those coherent structures, across the growing orders of magnitude, and making them scale invariant, up to a certain precision.

Therefore, we have a subset of IL, denoted A*, constituted by those expressions in IL that are, somehow, pragmatically distinguished as relevant to the description of the stated experimental outcome. And we have a set, $\Delta$, of formal clauses constraining the dispositions of the idealized users of IL to pragmatically valuate the expressions in A* that are to be parsed as formulas [98,99].

Assuming these restrained dispositions to be coherent, it follows that the structure of a unique, locally finite polyadic Boolean algebra of countable infinite degree, denoted $(A,I,S,\exists)$, is induced upon A* [98,99]. Here, 'A' stands for a Boolean algebra, 'I' denotes a countable infinite set, S is a function from transformations on I to Boolean endomorphisms on A, and $\exists$ is a function from the set of all subsets of I to quantifiers on A; S and $\exists$ are assumed to be compatible in a precise sense [34,36]. As to the Boolean algebra that can be obtained by the restriction of $(A,I,S,\exists)$ to closed formulas, it will be denoted B*.

Having set the adequate context, we are ready to present our first main result:

**Theorem 1.** *Let $u_n(x)$ be a convergent power series, defined exactly as above. If there exists a law of action, or operation on $Z_p[x]$ having the p-adic valuation as its* *parameter, and if it is subjected to the proper set of formal constraints, Δ, then, the result of that operation upon $u_n(x)$ is a uniquely defined element of $Z_p[x]$,*

$$u_{n+1}(x) \ = \ \Psi(u_n(x)) \ .$$

***Proof of theorem 1.*** We start with the observation that, as a convergent power series, $u_n(x)$ is also a continuous function from a subset of $p^n Z_p$ into $p^n Z_p$ (see [77], chap. IV.2):

$$u_n(x) \ \in \ C(p^n Z_p \to p^n Z_p)$$

Therefore, $u_n(x)$ is a continuous function from a zero-dimensional, totally disconnected compact Hausdorff space, which is to say, from a Stone (or Boolean) space into itself (see [70] and [48], chap. II.4, or [77], chap. I.3, or [81], chaps. I.5 and I.18, or [37], sect. 17).

Thus, using the topological duality of Stone [95-97], denoted $Dual_S$, we can, and we do translate $u_n$ by an uniquely determined homomorphism (see [37], sect. 20, or [57], chap. 3), $v_n$, from a certain Boolean algebra, $B_n$, into itself:

$$Dual_S \ : \ p^n Z_p \ \to \ B_n$$
$$u_n \ \mapsto \ v_n$$

Let us concentrate on this Boolean algebra, $B_n$, as we know that it is isomorphic to the Lindenbaum-Tarski algebra, $LT_n$, of some theory (see [57], chap. 4.9, and also, [67] and [75], chapts. 6 and 7). And let us take $lt_n$ as the image of $v_n$, by this isomorphism, BL:

$$BL \ : \ B_n \cong LT_n$$

$$v_n \mapsto lt_n$$

Now, by the hypothesis that was made above, in the prelusion to the statement of the theorem, concerning the ideality of the users of the inner language of the relevant physical theory, we have that $LT_n$ is correctly interpretable in $B_*$, the Boolean algebra that was mentioned to be obtainable by restriction of a certain polyadic algebra, $(A,I,S,\exists)$, to closed formulas.

Thus, we have an interpretary homomorphism, Hom (see [75], chap. 7.8 and 7.9),

$$Hom \ : \ LT_n \to B_* \ ,$$

sending all the objects in the ideal of theorems of $LT_n$ into the ideal of theorems of $B_*$.

So, just as we can express $lt_n$ as a consequence of the axioms of the theory of $LT_n$, we can also use Hom to express $Hom(lt_n)$ as a consequence of the axioms of $B_*$. Therefore, we can express $Hom(lt_n)$ as a consequence of the axioms of $(A,I,S,\exists)$.

Synthesizing all that was done so far, following the just exposed line of reasoning, we assert that, after the consecutive application of the topological duality of Stone, an isomorphism and an interpretary homomorphism, we get the expression of $u_n$ as a consequence of the axioms of $(A,I,S,\exists)$. In particular, this is how we link $u_n$ with a value taken by the map S, on the polyadic algebra:

$$S \ : \ I^I \to end_B A$$
$$i \ \mapsto \ Hom \ \{ \ BL \ [ \ Dual_S(u_n) \ ] \ \} \ ,$$

denoting the Boolean endomorphisms of A by $end_B A$, and the pertinent transformation on the set I by i.

It is quite clear that, if there exists an operation, $\Psi$ , on $Z_p[x]$, such that, in particular, its result is one of analytic extension into $p^{n-1}Z_p \setminus p^n Z_p$ , say,

$$u_{n+1} \; = \; \Psi(u_n) \; ,$$

then, we can apply this same procedure to $u_{n+1}$ in order to link it with a value taken by a map, $S^+$, on some other polyadic algebra, $(A^+, I^+, S^+, \exists^+)$, but not on $(A, I, S, \exists)$, as

$$u_{n+1} \; : \; p^{n-1}Z_p \; \to \; p^{n-1}Z_p \; .$$

However, given the formal constraints of the set $\Delta$, restraining the modalities for the operation $\Psi$ to obtain, the relation between these two polyadic algebras can not be an arbitrary one:

We remember that, by hypothesis, these formal constraints are defined, or thought to be stated by the idealized users of the inner language of the relevant physical theory [15], which is to say, those whose coherent interpretation was determinant to the induction of the polyadic structure [98,99].

Now, just as we may see that there is no proper topological additive subgroup between $p^{n-1}Z_p$ and $p^n Z_p$ , so will these idealized users be able to formalize the constraint on $\Psi$, to act in a manner that there will be no proper transformation subsystem of $(A^+, I^+, S^+)$ containing $(A, I, S)$, once that the choice of $I^+ \supset I$ is made, allowing $S^+$ to take a value that is linked to $u_{n+1}$ just like a certain value of S was previously linked to $u_n$.

Such a constraint, therefore, amounts to the requirement for $(A^+, I^+, S^+, \exists^+)$ to be a minimal [21] polyadic dilation of $(A, I, S, \exists)$ (see [35] or [36], chap. V) and, thereupon, given that the set I is infinite, for it to be the unique such one, to within equivalence and with respect to this passage, from the set I to the set $I^+$.

We proceed with the study of this passage, from $I$ to $I^+$. In order to do so, let us recover the set of convergent power series $(u_i)_{i=1,2,...,n}$ modeling the observed coherent structures across the corresponding orders of magnitude. We recall that these coherent structures were asserted to exhibit scale invariance, up to some precision, inside the window of observation; so, we formalize this assertion as a constraint in the set $\Delta$:

Let $B^-(1,1)$ denote the open disc with center 1 and radius $r < 1$.

Starting with a countable sequence of sides of zero, which is to say, with a countable sequence of multiplicative cosets of $B^-(1,1)$ (see [81], sect. 24), in a complete non-archimedean field including the p-adic numbers, $Q_p$, as a subfield, we take n of these sides of zero and we put them in a one-to-one correspondence with the power series $(u_i)_{i=1,2,...,n}$.

Next, we form the nonempty intersection, $U_n$, of each of these sides of zero with the codomain of the corresponding power series, and we choose the discs $B(y_n, r_n)$, open and closed, with center $y_n \in U_n$ and radius $r_n$ arbitrarily small, such that the ratios $r_{n+1}/r_n$ are proportional to the ratios between the orders of magnitude of the codomains of the corresponding power series.

Finally, let us be given a convenient, planar euclidean model of $Z_p$ (see [5,20,76] and [77], chap. I.2). We modelize the exhibited scale invariance as an homothety with ratio $r_{n+1}/r_n$ between the images that the orbit structures of $u_n$ and $u_{n+1}$, respectively, in $B(y_n, r_n)$ and $B(y_{n+1}, r_{n+1})$, both have in this euclidean model.

In these terms, the bottom-up constraint on the action, $\Psi$, upon $u_n$ will be the one that it be subjected to follow the projection of this homothety, towards the greater order of magnitude associated with $p^{n-1}Z_p$.

Once that this is granted, $\Psi(u_n)$ will obviously be defined in a set with an accumulation point and, therefore, uniquely defined in all of $U_{n+1}$ [58].

And thus, using, once again, the topological duality of Stone, this last statement can be translated to the definition of a unique homomorphism, from the Boolean algebra dual to $p^{n-1}Z_p$ into itself. Subsequently, after an isomorphism and an interpretary homomorphism, both conceived as above, we get the definition of a unique homomorphism from the Boolean algebra that results from the restriction of $(A^+, I^+, S^+, \exists^+)$ to closed formulas, into itself.

The passage from the set $I \subset I^+$ to the set $I^+$ can, thus, be restricted to the obtention of a precise value to be taken by $S^+$ :

$$\begin{aligned} S^+ \, &: \, (I^+)^{I^+} \, \to \, end_B A^+ \\ &\quad\; i^+ \, \mapsto \, Hom \, \{ \, BL \, [ \, Dual_S(u_{n+1}) \, ] \, \} \, , \end{aligned}$$

the remaining ambiguity in the polyadic dilation process being, thereby, removed. And the conclusion follows.

Focusing, momentarily, on the physical interpretation of the foregoing argument, we observe that our formal constraints can be thought to model the scale-dependent boundary conditions restraining the flow of an operator with the p-adic valuation as a parameter, transforming the topological, scale-dependent structure of the relevant physical fields.

## 3 - Existence.

We proceed with the analysis of the evolution, from one scale-dependent state representation to another scale-dependent state representation, of the same Sigma structure that was mentioned in section 1. Recall that this evolution is considered to underlie, and to determine the experimentally observed display of coherent structures, across the many orders of magnitude that are characteristic of a turbulent flow.

Subsequently to this observation, we want to concentrate ourselves on the connection between the models associated to the coherent structures detected at two, contiguous orders of magnitude, or scales, according to some p-adic valuation. We take the inference of these models for granted and we present them as the convergent power series written

$$u_n(x) = x^3 (1 + A_{n,1} x + A_{n,2} x^2 + ...)$$

$$u_{n+1}(x) = x^3 (1 + A_{n+1,1} x + A_{n+1,2} x^2 + ...),$$

with domains, respectively,

$$V_n = u_n^{-1} (p^n Z_p \setminus p^{n+1} Z_p)$$

$$V_{n+1} = u_{n+1}^{-1} (p^{n-1} Z_p \setminus p^n Z_p).$$

The observed coherent structures are supposed to exhibit scale invariance. We will modelize such a symmetry as being equivalent to the preservation of, at least, some of the topological orbit structure of these power series, taken as dynamical systems; and we will show that such a symmetry is nothing more than what is dictated on the terms of a law of action, making each of the given power series to be the analytic continuation of the other (see [24], pp. 76-80). The demonstration of the next theorem, stating this last assertion rigorously, is technically dependent on the use of an infraconnected set whose holes [26,77], judiciously distributed within a countable subsequence of mutually disjoint sides of zero [81], form themselves a countable subsequence that runs the T-filter [25,26] through which, using a result of Sarmant and Escassut [80], the analytic continuation is obtained.

**Theorem 2.** *Let $u_n$ and $u_{n+1}$ be two convergent power series defined exactly as above, and take an isometrical embedding of $Q_p$, the field of p-adic numbers, into* *an algebraically closed, spherically complete extension of itself admitting a non-countable residue class field; let $\Theta$ denote that extension, with valuation group equal to $[0, +\infty[$. If $u_n$ and $u_{n+1}$ are uniquely continuated to analitic elements, $w_n$ and $w_{n+1}$, belonging to the complete algebras of bounded analytic elements defined in the extensions, $\Omega_n$ and $\Omega_{n+1}$, respectively, of $V_n$ and $V_{n+1}$, in $\Theta$, then, there is a law of action making each of these analytic elements, $w_n$ and $w_{n+1}$, to be the analytic continuation of the other.*

***Proof of theorem 2.*** We shall proceed to show that, under the proper restrictions, the analytic element $w_{n+1}$ can be constructed as an analytic extension of $w_n$, but let us make two previous remarks, concerning the well founded character of our premises. First, as to the existence of the proposed field extension, $\Theta$, check [26], chapters 6 and 7, and [81], appendixes A.9 and A.10; second, as to the unique continuation of an analytic element, following an extension of the ground field, check also [26], page 87, theorem 15.9.

The following definitions and notation will hold, in the sequel; starting with the basics, we have:

All $r, r', r_1, r_2, r^*$ belong to $|\Theta| = \{ |x| : x \in \Theta \}$, with $|x|$ denoting the valuation of x in $\Theta$ that extends the one of $Q_p$.

$$p^{-n} < r < p^{-n+1} < r^* < 1$$

$$B^-(a,r) = \{ x \in \Theta : |x-a| < r \}$$

$$B^+(a,r) = \{ x \in \Theta : |x-a| \leq r \}$$

$$\Gamma(a,r_1,r_2) = \{ x \in \Theta : r_1 < |x-a| < r_2 \}$$

D will denote a bounded infraconnected [26,77] set, including the disc $B^+(0,r)$ and verifying two further conditions, to be specified ahead.

$H_B(D)$ denotes the complete algebra of bounded rational functions with no poles in the set D.

$I_0(B^+(0,r))$ denotes the ideal of those $y \in H_B(D)$ such that, for every $x \in B^+(0,r)$, $y(x) = 0$.

$\delta(0,B^-(a,r))$ will denote the distance of the disc $B^-(a,r)$ to the origin, 0.

Let r'>r, r' variable. We define a decreasing filter on D, with center $\xi \in B^+(0,r)$ such that $|\xi| = r$, and diameter r, to be one that admits for base the family of sets $\Gamma(\xi,r,r') \cap D$. Let us denote it F. We call $B^+(0,r)$ the beach of this filter. Observe that $B^+(\xi,r) = B^+(0,r)$, trivially.

Let $(r_n)_{n \in N}$ be a decreasing sequence such that $\lim_{n \to \infty} r_n = r$. We define the sequence of sets $\Gamma(\xi,r,r_n) \cap D$ to be a canonical base for F.

Having laid down the basic setting on which to work, we proceed to endow $\Theta \setminus \{0\}$ with an equivalence relation: given $x,x' \in \Theta \setminus \{0\}$, we say that x and x' are on the same equivalence class if 0 is not in the smallest disc containing x and x'.

These equivalence classes are nothing more than the already introduced sides of zero, and there are infinitely many of them; in fact, for each x in the ground field, we may write them as $xB^-(1,1) = B^-(x, |x|)$ (see [81], section 24). We shall take one countable subsequence, out of their set, in order to provide for the announced further qualifications on the set D:

Let us form the intersection of $B^-(x, |x|) \subset B^+(0,r^*)$ with $B^+(0,r^*)$, assuming that $|x|$ is the greatest value for which this inclusion of discs holds. We define the infraconnected set D as the following union of discs: $D = B^+(0,r) \cup B^-(x, |x|)$. Let $D^+$ stand for the closure of D.

We know that $B^+(0,r^*) \setminus D^+$ admits a unique partition ([26], p. 12, lemma 2.1) of the form $(T_n)_{n \in N} = (B^-(a_n, r_n))_{n \in N}$, whose elements are called the holes of D.

Now, we shall use the above mentioned countable subsequence of sides of zero (that we assume mutually disjoint) in order to define an infinite subsequence, $(T_m)_{m \in N}$, of holes of D. For every $n, m \in N$, we subject the holes in this infinite subsequence to the following specifications:

i) If $n \neq m$, then, $T_n$ and $T_m$ belong to different sides of 0.

ii) $r_{m+1} < r_m$ and $\lim_{m \to \infty} r_m = r$.

iii) $r_m = \delta(a_m, D) = \delta(\xi, T_m)$.

iv) Given the filter F on D, for some n on, each element of one of its canonical bases includes at most one element of the subsequence of holes $(T_m)_{m \in N}$.

Obeying these specifications, we have that the subsequence of holes $(T_m)_{m \in N}$ that we have just introduced is, by definition (see [26], pp. 21-22), a decreasing distances holes sequence that is simple and well pierced, has its center at the point $\xi$ and diameter r, and that, as such, runs a T-filter with the same characteristics, having the disc $B^+(0,r)$ as its beach (see [25] and [26], chapters 35 and 36, theorem 36.1).

It is clear that, besides from belonging to $H_B(B^+(0,r))$, the analytic element $w_n$ has its range included in $B^+(0,r)$. It follows, then, that, through this T-filter that we have managed to build, $w_n$ has analytic continuation, y, to analytic elements in $H_B(D)$ (see [80], theorem 4).

Let us restrict $w_{n+1}$ to $\Omega_{n+1} \cap D$. Then, $w_{n+1} \in H_B(D)$. We are going to use this fact in order to choose a particular element of $H_B(D)$.

Let $c, b \in V_{n+1} \cap D \subset \Omega_{n+1} \cap D$, and let $\tau, \tau' \in [0, \varepsilon]$, for $\varepsilon > 0$, small enough.

We take $B^-(c, \tau) \subset D \setminus B^+(0,r)$ and we form the sets

$$B^-(c, \tau) \cap V_{n+1} \cap D \quad \text{and} \quad B^-(b, \tau') \cap V_{n+1} \cap D = w_{n+1}(B^-(c, \tau) \cap V_{n+1} \cap D).$$

Next, we define the following parametrization, $\eta(\tau)$, on their cartesian product:

$$\eta: [0, \varepsilon] \to (B^-(c, \tau) \cap V_{n+1} \cap D) \times (B^-(b, \tau') \cap V_{n+1} \cap D)$$

$$\tau \mapsto (\eta_1(\tau), \eta_2(\tau)) = (x, w_{n+1}(x))$$

with $x \in B^-(c, \tau) \cap V_{n+1} \cap D$.

Let $y' \in I_0(B^+(0,r))$ be such that, for $\tau \in [0,\varepsilon]$, $y'(B^-(c,\tau)) \neq 0$.

We know that the infinitely many analytic elements that extend $w_n$ in $H_B(D)$, through our T-filter, are all of the form (see [80], theorem 4)

$$y + I_0(B^+(0,r)) ,$$

thus, so is

$$y_0 \;=\; y \;+\; \Big(\; [\eta_2(\tau) - y(\eta_1(\tau))] \;/\; [y'(\eta_1(\tau))] \;\Big) \; y' \;\; ,$$

along $\eta(\tau)$.

Thus, besides from being an analytic continuation of $w_n$, we have that $y_0$ agrees with $w_{n+1}$ in a set with an accumulation point. And this finishes the proof, as it obviously implies that $y_0$ agrees with $w_{n+1}$ at any point in the intersection of their domains of definition (see [58], § 2).

We observe that, from a strictly mathematical point of view, this second result (theorem 2) seems to supersede the previous one (theorem 1). However, from a physicist's perspective, the need for the result that is presented with theorem 1 prevails. In fact, if we try to model any well defined evolution of the topological structure of the relevant physical fields, by resort to an unique analytical extension through T-filters, we are confronted with the problem of the correspondence between the relevant boundary conditions and the criteria, which is to say, the parametrization that was adopted to choose one such analytical extension from an infinite set of possible ones; the formal constraints of theorem 1, to be stated in the inner language of the relevant physical theory, will, eventually, establish the bridge that closes that gap.

## 4 - Factorization.

The law of action that we have established can be thought to induce an arbitrary evolution of orbit structure in a set of formal power series. Now, we want to qualify such an evolution, in order to restrict ourselves to the analysis of paths that are compatible with the requirement for that action to be uniquely determined, once that it is subjected to the proper formal constraints.

The first step, in this direction, was taken by the imposition for the variation of structure to be minimal, but it is not enough, since we are still left with an evolution that proceeds, apparently, in a discrete way. And the case gets worse, on noticing that it was quite arbitrarily that we constructed the analytic extension through a T-filter to obtain between analytic elements defined at some contiguous scales. Indeed, there was nothing to impose us that the analytic extension could not be obtained connecting scales far apart from each other. Yet, such an apparently discrete evolution stands in contradiction, first, against the observed temporal continuity of generic turbulence phenomena, like the growth of coherent structures, for instance, and second, against the sustainable philosophical assertion that there are no unnatural gaps or boundaries in the universe for mathematical discourse to modelize such growth of coherent structures, for the same instance [2,14,62].

Having prompted these considerations upon our work, we proceed to set the context in which it will be possible for us to properly qualify the evolution that is induced by the action on $Z_p[x]$. Observe that this is how we intend to model the evolution, between scale-dependent state representations, of the yet undefined Sigma structures that we assume to underly any turbulence phenomena.

So, let us recall that our modeling procedure implies that we may interpret these scale-dependent state representations as p-adic dynamical systems; we may observe that the study of dynamical systems over p-adic fields has its precedents in the literature [5,30,47,63,105]. In fact, it has already been shown that it naturally leads to the study of formal groups over the integer ring of some finite extension of a p-adic field [63].

Reflecting on this background of ideas, our attention is driven by the works of Honda [41-43], and others [23,39], who used power series appropriate (generalized hypergeometric ones, for instance) in order to obtain the formal groups of some elliptic curves. Let us by-pass the technicalities involved, as we wish to retain only the idea of associating an elliptic curve, with a definite expansion near the origin (the formal group) [51,82,84], to each of our chosen, convergent power series.

This later option will be made at an expense, reducing the set of the admissible power series, but, as we further restrict ourselves to elliptic curves over the rationals and with semi-stable reduction [84], we get modularity [104,106] and rigid analytic structure [11,103], which is just what is needed so that we can prove that the established action exhibits a unique factorization property, along morphisms in the category of rigid analytic varieties. This is our third, main mathematical result; thereby, along the lines of its proof, the context will be set for the announced qualification on the action to be translated into an experimentally testable proposition, in section 6.

**Theorem 3.** *Let Y be the maximal ideal of a complete local field K with integer ring J, and let $u_1$ and $u_2$ be two convergent power series defined in Y, with coefficients in J. If there is an operation $\Psi$, such that $\Psi(u_1) = u_2$, then, $\Psi$ admits a non-trivial factorization along morphisms in the category of rigid analytic varieties. Moreover, this factorization is unique.*

*Proof of theorem 3.* Let C-VAR denote the category of rigid analytic varieties over a field K (see [11], chap. 9, or [102]).

We take K to be a field with a complete non-trivial non-archimedean valuation, algebraically closed and with residue characteristic $\neq 2$.

Let $\mathbb{Q}$ denote the field of the rational numbers. We assume that $\mathbb{Q} \subset K$. we also assume that K is the completion of an algebraic closure of a field, $\mathbb{Q}_p$, obtained by completion of $\mathbb{Q}$ with respect to a fixed p-adic valuation. The prime number p shall be determined afterwards.

Suppose that the power series $u_1$ and $u_2$ can be written in the form

$$u_1(x) = x^3 ( 1 + A_{1,1}x + A_{1,2} x^2 + \ldots )$$

$$u_2(x) = x^3 ( 1 + A_{2,1}x + A_{2,2} x^2 + \ldots ),$$

as power series whose coefficients

$$A_{i,j} \quad , \quad \text{with } i = 1,2 \text{ and } j = 1,2,3,\ldots ,$$

belong to the polynomial rational integer ring in the five rational integer variables

$$a_{ik} \quad , \quad \text{with } i = 1,2 \text{ and } k = 1,2,3,4,6 .$$

Then, $u_1$ and $u_2$ can be interpreted as those power series that, formally, give us the addition laws of two elliptic curves over $\mathbb{Q}$ (see [84], pp. 110-115, or [103], pp. 179-183); let $E_1$ and $E_2$, respectively, stand for those two such curves, with Weierstrass models defined

$$E_1 : \quad y^2 + a_{11}xy + a_{13}y = x^3 + a_{12}x^2 + a_{14}x + a_{16}$$

$$E_2 : \quad y^2 + a_{21}xy + a_{23}y = x^3 + a_{22}x^2 + a_{24}x + a_{26}$$

We apply Tate's algorithm to these equations (see [19], pp. 62-68, or [85], pp. 361-379) and we obtain the global minimal model of each curve. We also obtain, thereby, the reduction type of both $E_1$ and $E_2$, place by place, at each and every p. Consequently, we have all that we need to define the local L factors of two formal Euler products from which, in turn, we deduce the global L series of both curves, denoted, respectively (see [18], pp. 229-232, or [51], chap. X, or [55], chap. II, or [103], pp. 195-199):

$$L(s,E_1) \quad \text{and} \quad L(s,E_2) \,.$$

We now assume both $E_1$ and $E_2$ to be semi-stable elliptic curves, having bad reduction in at least one common place. Let p be the prime number for that place. It is this number, p, that we use to define the field underlying C-VAR.

Now, if $E_1$ and $E_2$ are semi-stable elliptic curves over Q, then, they are modular [104,106], which is to say, they both have modular parametrizations of levels $N_1$ and $N_2$, respectively, with $N_1$ and $N_2$ being identified with their conductors, respectively (see [18], p. 232, or [51], pp. 390-392).

Thus, we have that, in the vector space of cusp newforms of weight 2 and level M ($M=N_1,N_2$), denoted

$$S_2(\Gamma_0(M))^{new} \,,$$

where $\Gamma_0(M)$ stands for the adequate congruence subgroup of level M (with respect to the full modular group), there exist two cusp newforms with integer Fourier coefficients (see [51], p.374, theorem 11.74), denoted $f_1(z)$ and $f_2(z)$, and defined by the equalities

$$f_1(Z) = \sum_{n=1}^{\infty} a_1(n)e^{2\pi i nz} \in S_2(\Gamma_0(N_1))^{new}$$

$$f_2(Z) = \sum_{n=1}^{\infty} a_2(n)e^{2\pi i nz} \in S_2(\Gamma_0(N_2))^{new}$$

and that are such that

$$L(s,E_1) = L(z,f_1) \quad \text{and} \quad L(s,E_2) = L(z,f_2).$$

We proceed by giving to each global minimal model its Legendre form, and we calculate the absolute j-invariant for each curve [45].

Let $j_1$ and $j_2$ denote the absolute j-invariants of the elliptic curves $E_1$ and $E_2$, respectively, and let A stand for the Zariski-open subvariety obtained from affine one-dimensional space by removing the origin.

For each absolute j-invariant, say $j_*$, the correspondence given by

$$j_* \rightarrow q \quad,$$

and defined by the series

$$q = \sum_{n=1}^{\infty} h_n (j_*)^{-n}$$

with $j_*, q \in K$, $|h_1| = 1$, and the $h_n$ power limited, establishes a bijection between elliptic curves such that $|j_*| > 1$, on the one side, and one-dimensional analytic tori, denoted $A/(q)$, on the other side (see [11], pp. 414-415, or [103], p.190, theorem 5).

But, by assumption, both $E_1$ and $E_2$ are semi-stable elliptic curves having bad reduction at some common place, or prime. Thus, we have the inequalities

$$|j_1| > 1 \quad \text{and} \quad |j_2| > 1 \quad,$$

and thereby, there are two well determined one-dimensional analytic tori over K, denoted $A/(q_1)$ and $A/(q_2)$, corresponding to the elliptic curves $E_1$ and $E_2$, respectively.

These tori are rigid analytic varieties over K (see [11], pp. 364-365). Given that they have an underlying affinoid structure, and that there exist fibre products in the category of K-affinoid varieties, it follows that, for some object $S \in$ C-VAR, a fibre product, denoted

$$A/(q_1) \times_S A/(q_2) \quad,$$

can be constructed in the category of rigid analytic varieties over K (see [11], pp. 365-368). Such a fibre product defines a rigid analytic variety over K (and over S). We are going to give it an absolutely unique canonical structure, choosing a specific product diagram (see [8], p. 120):

We adjoin to $A/(q_1)$ the structure, $(f_1)$, of an element, $f_1$, of the vector space of cusp newforms of weight 2 and level $N_1$; and we proceed analogously with respect to $A/(q_2)$. Next, we adjoin to $A/(q_1) \times_S A/(q_2)$ the structure, $[(f_1,f_2)]$, of an element, $(f_1,f_2)$, of the vector space that is the direct Hilbert sum written

$$S_2(\Gamma_0(N_1))^{new} \oplus S_2(\Gamma_0(N_2))^{new} \quad.$$

And we can conclude: the operation $\Psi$, such that $\Psi(u_1) = u_2$, if it exists, has a unique factorization (relatively to some object $S \in$ C-VAR) along morphisms in C-VAR, that is constructed with the coordinate projections of the specified fibre product diagram:

$$u_1 \to A/(q_1) \xrightarrow{proj_1^{-1}} A/(q_1) \times_S A/(q_2) \xrightarrow{proj_2} A/(q_2) \to u_2 \quad.$$

**5 - Match between L-functions.**

In this section, we take a closer look at the connection between our research and the theory of qualitative dynamics. In particular, we delineate an approximation method that will be useful when trying to establish a protocol for the analysis of experimental data, testing the proposition that we will put forward, in section 6. We resort to analytic Number Theory, as developed by Knopfmacher [52-54], in order to achieve such a tool for future use.

Let us start with the observation that, generally speaking, it is quite reasonable to approximate the orbit structure of a dynamical system through the study of its set of periodic points, with period $m = 1,2,...,$ and through the construction of an adequate zeta function, gathering the data thereby obtained. Such an approach, which is naturally induced by the consideration of zeta functions counting the number of fixed points of the Frobenius morphism of an algebraic variety, is a clear suggestion of the relevance that pure Number Theory may have to our purposes (see [89], pp. 764-766, and [78], pp. 1-9).

In fact, under certain restrictive conditions, namely, the validity of Axiom A, there results an analytic treatment from which it follows the importance of the prime number theorem in this field of research (see [71]).

Thus, we acknowledge that, if there exists an L function of a strictly arithmetical nature to be associated with $L(s,E_1)$ and $L(z,f_1)$, for instance, then, that function must exhibit properties close to the ones that the above-mentioned zeta function may show, as long as we consider the original power series to be interpretable as dynamical systems.

So, let us reconsider $L(s,f_1)$:

$$L(s,f_1) \;=\; L\left(s, \sum_{n=1}^{\infty} a_1(n)e^{2\pi i n z}\right) \;=\; \sum_{n=1}^{\infty} a_1(n) n^{-s}.$$

Let $n_1, n_2, n^*$ denote some rational integers, with $n^* > 0$, $n^*$ arbitrarily big. We assume $p^*$ to denote a prime number such that $p^* > n^*$, but also, that $p^*$ is minimal with respect to $n^*$, in a sense to be specified ahead.

We define $G_Z[p^*]$ to be the arithmetical semigroup (commutative semigroup with identity, 1, equipped with a norm denoted $|\;|_G$) of all positive integers that are co-prime to $p^*$, and we let $\equiv_{p^*}$ stand for the congruence modulo $p^*$ equivalence relation.

We proceed to define $\chi_{p^*}(n)$ as a non-principal Dirichlet character modulo $p^*$ [3]. Excluding the case $n = 1$, for which it is $\chi_{p^*}(1) = 1$, we do this at the expense of the next $n^*-1$ coefficients $a_1(n)$ in the following way:

First, we normalize these coefficients to unit:

$$a_1(n) \;\rightarrow\; a_1(n) \;/\; \sum_{m=2}^{n^*} a_1(m) \;=\; n_1/n_2 \;=\; (n_1/n_2)(n).$$

Second, we apply the normalized coefficients into the multiplicative group of all complex numbers of absolute value 1:

$$(n_1/n_2)(n) \;\rightarrow\; \exp(i\pi n_1/n_2) \;=\; \chi_{p^*}(n).$$

The prime number $p^*$ is minimal with respect to $n^*$ in the sense that it is taken to be the smallest prime for which this definition obtains.

As it is, we then have that the ordered pair

$$( G_Z[p^*] , \equiv_{p^*} )$$

defines a number theoretical structure called arithmetical formation (see [28]), whose L function we write as (see [52,53], or [54], chap.9)

$$L_{G_Z[p^*]}(s,\chi_{p^*}) = \sum_{n \in G_z[p^*]} \chi_{p^*}(a_1(n)) \, n^{-s} \, ,$$

and we obtain that, in particular,

$$\sum_{n=2}^{n^*} a_1(n) \, n^{-s} = (n_2/\pi) \sum_{n=2}^{n^*} \cos^{-1} [ \, Re \, (\chi_{p^*}(a_1(n))) \, ] \, n^{-s} \, .$$

We recall that $n^* < \infty$ can be as big as we want. Besides, we know that (see [4], p.134, theorem 6.17, or [83], p. 94, theorem 5)

$$a_1(n) = O(n) \, ,$$

thus, we are able to control the error made when using

$$L_{G_Z[p^*]}(s,\chi_{p^*})$$

to estimate $L(s,f_1)$, in the region where it is known that both series are absolutely convergent ($Re(s)>2$).

The arithmetical formation $(G_Z[p^*] ,\equiv_{p^*})$ has exactly $\varphi(p^*)$ equivalence classes (with $\varphi$ denoting the Euler totient function). Let H denote one of its classes and define

$$N_H(x) = \# \{ \, y \in H : \, |y|_G < x \, , \, x > 0 \, , \, x \text{ real} \, \}$$

$$\pi_H(x) = \# \{ p \in H : |p|_G < x , x > 0 , x \text{ real} , p \text{ prime} \} .$$

Then, it follows that (Axiom A* for this formation)

$$N_H(x) = (p^*)^{-1} x + O(1) ,$$

when $x \to \infty$, and that (abstract prime number theorem for this formation)

$$\pi_H(x) \sim x [ \varphi(p^*) \log x ]^{-1} ,$$

when $x \to \infty$. That is, the arithmetical L function that we used to approximate $L(s,f_1)$ is, indeed, connected to an arithmetical structure whose analytic and number theoretical properties are analogous to the ones encountered in the studies of differentiable dynamical systems obeying Axiom A (Axiom A diffeomorphisms), using zeta functions (compare [89], pp. 764-766, and [71] with [54], pp. 264-285).

**6 - Match with experimental data.**

In this section, our goal will be to use the previously exposed theory, specifically, in sections 2 and 4, in order to infer from it a certain set of experimentally testable propositions, qualifying the general thesis that the evolution between scale-dependent state representations (which is to say, in particular, between the topological structures in the phase portraits of the adequate power series) of the Sigma structures underlying turbulence is one that unfolds along geodesic paths.

To achieve this goal, we will do analogously to what we did in section 2, translating the physical requirement for the action to be minimal into rigorous mathematical terms. We recall that the expression that was, then, proposed was the demand for a certain polyadic dilation to be minimal.

Now, we want to depart from the local character of that former expression, so that we can cover situations where the choice among the available paths, between state representations, results in the choice that minimizes variation of structure.

The concrete structure whose variation we minimize here is that one, already introduced in section 4, of cusp newforms of weight 2 and level arbitrary. The consequence of this constraint, on the evolution of orbit structure inside our set of power series will, therefore, be obtained through modular elliptic curves. So, let us proceed by the setting of an elliptic surface [85,93,94].

**6.1 - The framework.**

We recover the levels of modular parametrization that were used in section 4, $N_1$ and $N_2$, but we now impose that

$$N_1 = N_2 = N,$$

and just to simplify things, we also assume that the vector space of cusp newforms of weight 2 and level N has dimension one:

$$dim\ (\ S_2(\Gamma_o(N))^{new}\ )\ =\ 1\ .$$

We use the linear combination

$$(1-t)f_1(z)\ +\ t f_2(z)\ ,$$

with $t \in [0,1] \cap \mathbf{Q}$, to define a family of newforms on this space (after some denominator clearance, if necessary). Let $P^1(\mathbf{Q})$ denote one-dimensional rational projective space. We define a bijection $\gamma$, such that

$$\gamma : [0,1] \cap \mathbf{Q} \to P^1(\mathbf{Q})$$

$$\gamma(t) = (c,t) \quad , \quad c \neq 0 .$$

Now, let F be a two-dimensional projective variety, and let $\{\Pi_i\}_{i=1,2,3,4,6}$ be a set of five morphisms defined

$$\Pi_i \cong A_i^{-1}(c,t) : F \to P^1(\mathbf{Q}) ,$$

and such that, for all but finitely many t, the fiber, $E_t$, defined by the equation

$$y^2 + A_1(c,t) xy + A_3(c,t) y = x^3 + A_2(c,t) x^2 + A_4(c,t) x + A_6(c,t) ,$$

with

$$A_i(c,0) = a_{1i} \quad \text{and} \quad A_i(c,1) = a_{2i} ,$$

is a non-singular curve of genus 1, which is to say, an elliptic curve.

Then, we acknowledge that our family of cusp newforms induces a fibration which, in turn, allows for the definition of an elliptic surface over $P^1(\mathbf{Q})$ (see [85], pp. 202-203). And we see that, along this fibration, just as $E_1$ is deformed onto $E_2$, so the orbit structure of $u_1$ is transformed onto the orbit structure of $u_2$.

## 6.2 - The first trial.

We now identify the ground field with $Q_p$ and assume that its integer ring, $Z_p$, has a well determined fixed planar euclidean model in the unit disc $U \subset R^2$ (where $R^2$ stands for the two dimensional vector space over the field of the real numbers) [76,77]. And we identify U with an arbitrary cross section of a material cylindric tube having an internal fixed grid that is normal and symmetric with respect to the axis of the cylinder.

We suppose that there is a fluid flowing in the tube and that, behind the grid, a regular sucession of vortices known as the von Kármán vortex street forms (see [32], pp. 1-13 and [60], pp. 240-241, for a qualitative description, or [79], pp. 133-138, for a quantitative treatment).

We take two cross sections of the tube, separated from each other by a small distance, both in the region of relaxation of turbulence. Each of these two sections has some vorticity vectors applied on the discs that they define (see [73], pp. 1.5-1.8, or [79], chapter 1). We project those vectors onto the discs. Now, these projections define two vorticity fields whose field lines we are going to interpret as being representatives of the images of the orbit structure, or phase portrait of some power series, $u_1$ and $u_2$, on the above-mentioned planar euclidean model.

Assuming these power series to be quite effective models of the topological structure of the vorticity field, it is obvious, on a physical ground, that there must exist an operation $\Psi$, such that $\Psi(u_1) = u_2$. Thus, the possibility exists, for there to be an elliptic surface associated to $\Psi$. Such an elliptic surface, however, can not be an arbitrary one, if its fibers, between $E_1$ and $E_2$, are to be constrained to reflect, somehow, the physical restrictions on the evolution of phase portraits that are typical of the chosen example (entropy growth, energy transfer from the turbulence to the fluid) (see [60], chap. 14). Thus, in particular, the choice of a map like the above one, from $[0,1] \cap Q$ to $S_2(\Gamma_o(N))^{new}$, must be clearly justified.

Consequently, we now choose a specific elliptic surface for this problem.

Let us maintain the assumption that $E_1$ and $E_2$ have modular parametrizations of the same level N.

It is known that the vector space $S_2(\Gamma_o(N))^{new}$ has a canonical Hilbert space structure on it (see [107], pp. 239-240 and part 2, and [51], pp. 231-255). Therefore, we can use its inner product, the Petersson inner product, ( , ), defined

$$(f_1, f_2) = \iint_{R_o} v^2 f_1(\mu + i\nu) f_2^*(\mu + i\nu) \, v^{-2} d\mu d\nu \,,$$

where $R_o$ denotes a fundamental domain for $\Gamma_o(N)$, in the complex upper semi-plane, the symbol * stands for the complex conjugate, and $v^{-2}d\mu d\nu$ is the appropriate invariant measure, in order to establish a metric, d, defined by the equality (see [31], pp. 83-84)

$$d[f_1, f_2] = (f_1-f_2, f_1-f_2)^{1/2} \,.$$

Naturally, if the above mentioned cross sections are made close enough to each other, along the turbulence, then, we expect the orbit structures of $u_1$ and $u_2$ to be topologically conjugate to each other [1,89]. Therefore, the continuity of the sesquilinear form [12] defined by the Petersson inner product, which mirrors quite well this expectation, seems to point the obvious choice to be made, always respecting the extremum principle that must be observed, for the action $\Psi$ to be qualified as physically meaningful:

The elliptic surface that we seek for, the one to which we give physical significance, is precisely the one for which the fibration splits, under the natural order relation in $[0,1] \cap \mathbf{Q}$, in such a way that both the variation of distances between the newforms associated to $E_1$ and $E_2$ and the newform associated to the current fiber, $E_t$, get to be extreme, step by step.

We call such a surface, denoted β, the geodesic elliptic surface. Observe that the assertion of its choice constitutes the first proposition that we submit to the trial of experience, notwithstanding some simplifications that were implicitly made in the argument that led us to it; these simplifications, namely, the omissions both of obstructions, or singularities, and of boundary conditions, will be superseded ahead, in the second trial (6.3).

Meanwhile, observe further that, following the argument that was exposed in section 5, it became reasonable to associate the arithmetical formation that was, then, introduced, with an auxiliary, abstract dynamical system, the L-function of the former naturally counting the number of isolated periodic points of the later. Therefore, if we now 'translate' the newform associated to $E_2$ to become the origin of its own vector space, then, the evolution along β, the geodesic elliptic surface, will be such that the auxiliary, abstract dynamical system associated to the current fiber, $E_t$, via the match of L-functions will automatically evolve in the way of maximizing its numerical stability [6], which is indeed quite an interesting circumstance to realize.

**6.3 - The second trial.**

Let us now consider the application of our results to the physics of plasma confinement [38].

There is some experimental evidence available, showing that the development of turbulence in toroidal plasmas can lead to the formation of coherent vortical structures like those in turbulent fluid flow [9,74]. However, for reasons that will become apparent, we prefer to concentrate on the magnetic induction field as the object of our modeling procedure. So, let us keep $Q_p$ as the ground field. We also assume $Z_p$ to have some fixed planar euclidean model in the unit disc $U \subset R^2$, as before, but, now, we identify this disc with an arbitrary poloidal section taken from the interior of a magnetic confinement chamber of the tokamak type (see [38], pp. 7-18). Thus, we see U as a poloidal section of a torus.

Suppose that there is a plasma column in the chamber, and that its dynamics is dominated by some precursory modes, prior to collapse (see [46]). Suppose also that the plasma column collapses indeed.

Take two poloidal sections, made in the same place but separated by a reasonable lapse of time, before and after collapse.

The magnetic induction field that is responsable for the confinement of the plasma column has some of its vectors applied on both poloidal sections. Let us project these vectors onto the discs defined by the sections; what we get, then, are two magnetic induction fields on the discs (see [86]).

We consider the field lines of these two late fields and we interpret them as representatives of the images of the orbit structures of some power series, $u_1$ and $u_2$, in the planar euclidean model of $Z_p$ that was mentioned above (see [76,77]).

Note that it is the temporal evolution of the topological properties of these images of orbit structure that governs, and dictates the quality of confinement: a single stable magnetic toroidal surface, acting as a transport barrier, can be enough to ensure the confinement of the plasma [10,29,33,38,44,61,68].

Here again, it is obvious, on a physical ground, that, if the chosen models are effective, then, there must exist an operation $\Psi$, such that $\Psi(u_1) = u_2$. And that there are restrictions of a physical nature that must be respected by $\Psi$ (growth of entropy, conversion of magnetic energy into kinetic and thermal energies), much in the same way as it was the case with the von Kármán vortex street.

The novelty, with respect to the previous trial, comes from the fact that now, necessarily, $\Psi$ does not preserve the topology (in an extension that we know to be proportional to the loss of magnetic energy): due to collapse, there can be no topological conjugacy between $u_1$ and $u_2$ (see [10], chap. 6, or [38], chap. 9; see also [16,46,61]).

The geodesic elliptic surface for this trial is now the underlying path for the evolution of some magnetic induction field lines in three distinct, sucessive regimes (the precursory, magnetic trigger and relaxation modes, or phases, respectively, before, during and after collapse). Let us see how can all this information be encoded in it.

We take it as self-evident that, being dependent on some, contingent boundary conditions, such a precise disruption of topological orbit structure is merely acidental. But, the matters must be different, we sustain, in what concerns to the underlying Sigma structure:

In fact, since we have imposed the reference of our Sigma structure to remain undetermined, it follows that its correspondent evolution of state can not depend on the detailed physics of the material support that unveils its presence, on developing turbulence.

These physical particularities, like general, contingent boundary conditions that they are, can only affect the representations that we manage to construct along the modeling procedure, as these later correspond to quantities that are measurable through some given, limited window of observation and, further, elaborated on the terms of some few, limited universes for mathematical discourse.

Now, taking a somewhat closer look to the modeling procedure, it is clear that the due step, for the inversion of experimental data to occur, is one that must result in the obtention of coefficients of certain power series and, subsequently, on the determination of the Weierstrass coefficients of certain elliptic curves [84,103]. Having obtained these elliptic curves, then, our analysis must unfold along two, parallel lines of thought:

On the one side, we resort to the proof of the Taniyama-Weil conjecture, by Wiles [106] and Taylor-Wiles [104], and we use a theorem of Carayol to connect it with the Eichler-Shimura theory [51], obtaining, thereby, a mathematical setting where the physicists request for an extremum of the action can be made compatible with the mathematicians demand for there to be some permanence of structure (the space $S_2(\Gamma_o(N))^{new}$ is closed under the operation $\Psi$, minimizing distances, just like in section 2, the minimal polyadic dilations were observed to preserve the polyadic structure).

On the other side, we recur to the isomorphism from semi-stable elliptic curves onto one-dimensional analytic tori [11], and we interpret these later as the local coordinate charts for the rigid analytic variety which is their fiber product and into which we frame the state representations defined by those former elliptic curves. Thereby, the connection between state representations that is assured to us by the action gets to be interpreted as the transition function between these local charts.

Observe that, within the framework that was set in this section (6.1), these two lines of thought are parallel to each other in the following sense: given a path of states, defined by the action on the scale-dependent state representation of a Sigma structure, in $S_2(\Gamma_o(N))^{new}$, there is a parametrization from this path onto an elliptic surface where the observable representations of those states may be identified with the appropriate fibers.

Quite obviously, if the above proposed interpretation is valid, then, for any fiber to be qualified as appropriate, beside from being good, it must correspond, neither to an unstable elliptic curve, nor to a stable one. In general, whenever the map from $S_2(\Gamma_o(N))^{new}$ takes one of these types of elliptic curves as its value, we will assume that the state at the source of that map has no observable representation at its target; and an observable discontinuity in the smooth evolution of topological orbit structure must follow.

Not quite so obviously, the same can be said, with respect to the causal effects that each singular set of boundary conditions must exert on the evolution of the physical fields. Once that a certain geodesic path is defined in $S_2(\Gamma_o(N))^{new}$, the possibility appears for there to be a set of positive measure in it, having as its elements states for which the representation is forbidden by the physical criterion that is dictated on the terms of some qualified boundary conditions. Consequently, we deny them any physical significance, whatsoever.

So, it may happen that, along some geodesic elliptic surfaces, there occur gaps between good fibers with local coordinate charts. And the magnitude of these gaps may be large enough so that, to the transition between the local charts that they separate, there corresponds some topological transition between orbit structures that implies the release of a measurable amount of energy. In particular, an amount of energy of the same order of magnitude of the one that is typical of collapse events, in toroidal plasmas [7,10,16,27,44,46,61]. This is the second proposition that we submit to the trial of experience.

As a corollary, it follows that, in this way, precursory modes get to be interpreted as a consequence of an approach to some gaps on a given elliptic surface (or the crossing of a segment of it, with a number of small gaps), and the collapse event turns out to be nothing more than the crossing of some large enough gap. Thus, the careful analysis of the geodesic β may allow for the transition between the precursory modes and the trigger modes to be determined, previously to the event of collapse.

## 7. Final considerations.

We find it worthwhile to emphasize, although briefly, the two following points:

First, that it should be clear that the fine tuning of the formal constraints acting, either as bottom-up, or as top-down constraints on the law of action that was studied is extendable to the modelization of whatever symmetry breaking one may wish to consider, as our restriction to exact scale invariance, in section 2, was made only to facilitate the exposition.

Second, that it is now possible to infer that we have achieved independence of the ground number field in which the parametrization of turbulences takes place: according to the exposition made in section 4, whenever three, consecutive semi-stable elliptic curves in β do not share one, common, bad reduction prime, then, it becomes necessary to change the ground number field; alternatively, whenever two, consecutive semi-stable elliptic curves share more than one, bad reduction prime, then, the choice between the corresponding ground number fields is arbitrary.

Our work was conducted upon the assumption that, underlying the mathematics of turbulence, there is some species of structure, Sigma, of which we don't manage to know more than a few transient models, or representations in the corresponding categories.

Thus, one question remains to be settled, that is, if there are reasons for the network of isomorphisms that we used to stop unfolding where we left, in order to fix Sigma's reference.

The fact is that we linked $\Psi$ with the transformation properties of, simultaneously, semi-stable elliptic curves, modular cusp newforms, one-dimensional analytic tori, analytic elements, Stone spaces, Boolean, Lindenbaum-Tarski and polyadic algebras! This fact may be thought as the establishment of a network of adjunctions between universes for mathematical discourse to conceptualize such a species of structure [13,64]. But, if this is not enough, from a mathematician's or a logician's perspective [59], to achieve referential determinacy, the fact also is, that, from a physicist's point of view, the degree of invariance that was hereby achieved [72], in the description of that, sought for, mathematical species of structure, by $\Psi$, seems quite suitable to modelize the evolution of structure in turbulence, in whatever form that it presents or develops itself in the world, which is to say, independently from the specific nature of the systems supporting it. And this is how close we can get to the starting point of our work.

**Acknowledgement.** The author acknowledges the word of encouragement of his father, late academician, major-general José Guilherme Calvão Borges (†2001), during the final stages of preparation of this manuscript.

# 8 - References.

**1.** Abraham, R., Marsden, J.E.: *Foundations of Mechanics*, 2nd ed., Addison-Wesley Publishing Company, Reading, Massachusetts, 1995.

**2.** Almog, J.: The Plenitude of Structure and Scarcity of Possibilities, The Journal of Philosophy **88**, 11, 620-622 (1991).

**3.** Apostol, T.M.: *Introduction to Analytic Number Theory* , UTM, Springer-Verlag, New York, 1995.

**4.** Apostol, T.M.: *Modular Functions and Dirichlet Series in Number Theory,* $2^{nd}$ ed., GTM vol. 41, Springer-Verlag, New York, 1997.

**5.** Arrowsmith, D.K., Vivaldi, F.: Geometry of p-adic Siegel discs, Physica **71**D, 222-236 (1994).

**6.** Artin, E., Mazur, B.: On Periodic Points, Ann. of Math. **81**, 82-99 (1965).

**7.** Avinash, K., Bulanov, S.V., Esirkepov, T., Kaw, P., Pegoraro, F., Sasorov, F., Sen, A.: Forced Magnetic Field Line Reconnection in Electron Magnetohydrodynamics, Phys. Plasmas **5**, 2849-2860 (1988).

**8.** Barr, M., Wells, C.: *Category Theory for Computing Science*, Prentice Hall, London, 1990.

**9.** Benkadda, S., Dudok de Wit, T., Verga, A., Sem, A.: ASDEX team and X. Garbet, Characterization of Coherent Structures in Tokamak Edge Turbulence, Phys. Rev. Lett. **73**, 25, 3403-3406 (1994).

**10.** Biskamp, D.: *Nonlinear Magnetohydrodynamics*, Cambridge University Press, Cambridge, 1997.

**11.** Bosch, S., Güntzer, U., Remmert, R.: *Non-Archimedean Analysis. A Systematic Approach to Rigid Analytic Geometry*, Grundlehren der mathematischen Wiessenschaften vol. 261, Springer-Verlag, Berlin, Heidelberg, 1984.

**12.** N. Bourbaki, *Espaces Vectoriels Topologiques*, Masson, Paris, 1981.

**13.** Bourbaki, N.: *Théorie des Ensembles*, Masson, Paris, 1990.

**14.** Bricker, P.: Plenitude of Possible Structures, The Journal of Philosophy **88**, 11, 607-619 (1991).

**15.** Bugajski, S.: What is Quantum Logic, Studia Logica **XLI**, 4, 311-316 (1982).

**16.** Bulanov, S.V., Echkina, E.Yn., Inovenkov, I.N., Pegoraro, F., and Pichushkin, V.V.: On the Structural Stability of Magnetic Configurations With Two Null Lines, Phys. Plasmas **6**, 802-815 (1999).

**17.** Chirikov, B.V.: Patterns in Chaos, in Artuso, R. et al. (eds.) *Chaos, Order and Patterns*, NATO ASI Series B: Physics Vol. 280, Plenum Press, 1991.

**18.** Cohen, H.: Elliptic Curves, in Waldschmidt, M. et al. (eds.) *From Number Theory to Physics*, Springer-Verlag, Berlin, Heidelberg, 1995.

**19.** Cremona, J.E.: *Algorithms for Modular Elliptic Curves*, 2$^{nd}$ ed., Cambridge University Press, Cambridge, 1997.

**20.** Cuoco, A.: Visualizing the p-adic Integers, Amer. Math. Monthly **98**, 355-364 (1991).

**21.** Daigneault, A., Monk, D.: Representation Theory for Polyadic Algebras, Fundamenta Math. **LII**, 151-176 (1963).

**22.** Davidson, D.: Reality Without Reference, Dialectica **31**, 3-4, 247-258 (1977).

**23.** Deninger, C., Nart, E.: Formal Groups and L-Series, Comment. Math. Helvetici **65**, 318-333 (1990).

**24.** Dennery, P., Krzywicki, A.: *Mathematics for Physicists*, Dover Publications, Inc., New York, 1996.

**25.** Escassut, A.: T-Filtres, Ensembles Analytiques et Transformation de Fourier p-Adique, Ann. Inst. Fourier, Grenoble **25**, 45-80 (1975).

**26.** Escassut, A.: *Analytic Elements in p-Adic Analysis*, World Scientific, Singapore, 1995.

**27.** Fischer, O., Cooper, W.A.: Topologic Study of a Magnetic Perturbation in a 3D Free Boundary Plasma Equilibrium, Plasma Phys. Control. Fusion **40**, 1269-1284 (1998).

**28.** Forman, W., Shapiro, H.: Abstract Prime Number Theorems, Commun. Pure Appl. Math. **7**, 587-619 (1954).

**29.** Frank, A.G.: Magnetic Reconnection and Current Sheet Formation in 3D Magnetic Configurations, Plasma Phys. Control. Fusion **41**, A687-A697 (1999).

**30.** Freund, P.G.O., Olson, M.: p-Adic Dynamical Systems, Nuclear Phys. **297**B, 86-102 (1988).

**31.** Fréchet, M.: *Les Espaces Abstraits*, Gauthiers-Villars et Cie, Paris, 1928.

**32.** Frish, U.: *Turbulence*, Cambridge University Press, Cambridge, 1998.

**33.** Greene, J.M.: Reconnection of Vorticity Lines and Magnetic Lines, Phys. Fluids **5**B, 2355-2362 (1993).

**34.** Halmos, P.R.: The Basic Concepts of Algebraic Logic, Amer. Math. Monthly **53**, 363-387 (1956).

**35.** Halmos, P.R.: Algebraic Logic, II. Homogeneous Locally Finite Polyadic Boolean Algebras of Infinite Degree, Fundamenta Math. **XLIII** (1956), 255-325 (1956).

**36.** Halmos, P.R.: *Algebraic Logic*, Chelsea Publishing Company, New York, 1962.

**37.** Halmos, P.R.: *Lectures on Boolean Algebras*, D. Van Nostrand Company, Inc., Princeton, New Jersey, 1963.

**38.** Hazeltine, R.D., Meiss, J.D.: *Plasma Confinement*, Addison-Wesley, New York, 1992.

**39.** Hill, W.L.: Formal Groups and Zeta-Functions of Elliptic Curves, Inventiones math. **12**, 321-336 (1971).

**40.** Holmes, P., Lumley, J.L., Berkooz, G.: *Turbulence, Coherent Structures, Dynamical Systems and Symmetry*, Cambridge University Press, Cambridge, 1998.

**41.** Honda, T.: Formal Groups and Zeta-Functions, Osaka J. Math. **5**, 199-213 (1968).

**42.** Honda, T.: On the Theory of Commutative Formal Groups, J. Math. Soc. Japan **22**, 213-246 (1970).

**43.** Honda, T.: Formal Groups Obtained From Generalized Hypergeometric Functions, Osaka J. Math. **9**, 447-462 (1972).

**44.** Hornig, G., Schindler, K.: Magnetic Topology and the Problem of its Invariant Definition, Phys. Plasmas **3**, 781-791 (1996).

**45.** Igusa, J.-I.: Fibre Systems of Jacobian Varieties (III. Fibre Systems of Elliptic Curves), Amer. J. Math. **81**, 453-476 (1959).

**46.** Itoh, S.-I., Itoh, K., Zuschi, H., Fukuyama, A.: Physics of Collapse Events in Toroidal Plasmas, Plasma Phys. Control. Fusion **40**, 879-929 (1998).

**47.** Jang, Y.:Non-Archimedean Quantum Mechanics, Thesis, Tohoku Math. Publ. **10** (1998).

**48.** Johnstone, P.T.: *Stone Spaces*, Cambridge stud. adv. math., vol. 3, Cambridge University Press, Cambridge, 1982.

**49.** Juarrero, A.:*Dynamics in Action. Intentional Behavior as a Complex System*, A Bradford Book, The MIT Press, Cambridge Massachusetts, 1999.

**50.** Kagan, Y.Y.: Seismicity: Turbulence of Solids, Nonlinear Sci. Today **2**, 1-13 (1992).

**51.** Knapp, A.W.: *Elliptic Curves*, Math. Notes vol. 40, Princeton University Press, Princeton, New Jersey, 1992.

**52.** Knopfmacher, J.: Arithmetical Properties of Finite Rings and Algebras, and Analytical Number Theory. IV: Relative Asymptotic Enumeration and L-Series, J. Reine Angew. Math. **270**, 97-114 (1974).

**53.** Knopfmacher, J.: Arithmetical Properties of Finite Rings and Algebras, and Analytic Number Theory. V: Categories and Relative Analytic Number Theory, J. Reine Angew. Math. **271**, 95-121 (1974).

**54.** Knopfmacher, J.: *Abstract Analytical Number Theory*, Dover, New York, 1990.

**55.** Koblitz, N.: *Introduction to Elliptic Curves and Modular Forms*, $2^{nd}$ ed., GTM vol. 97, Springer-Verlag, New York, 1993.

**56.** Koblitz, N.: *p-adic Numbers, p-adic Analysis, and Zeta-Functions*, $2^{nd}$ ed., GTM vol. 58, Springer-Verlag, New York, 1996.

**57.** Koppelberg, S.: Elementary Arithmetic, in Monk, J.D., Bonnet, R. (eds) *Handbook of Boolean Algebras*, Elsevier Science Publishers B.V., 1989.

**58.** Krasner, M.: Prolongement Analytique Uniforme et Multiforme dans les Corps Valués Complets, in Krasner, M. (ed) *Les Tendances Géométriques en Algèbre et Théorie des Nombres*, Éditions du CNRS, Paris, 1966.

**59.** Kripke, S.A.: Naming and Necessity, in Davidson, D., Harman, G. (eds) *Semantics of Natural Language*, Reidel, Dordrecht, 1972.

**60.** Kuethe, A.M., Schutzer, J.D.: *Foundations of Aerodynamics*, John Wiley & Sons, New York, 1957.

**61.** Kulsrud, R.M.: Magnetic Reconnection in a Magnetohydrodynamic Plasma, Phys. Plasmas **5**, 1599-1606 (1998).

**62.** Lewis, D.: *On the Plurality of Worlds*, Basil Blackwell Ltd., Oxford, 1986.

**63.** Lubin, J.: Nonarchimedean Dynamical Systems, Compositio Math. **94**, 321-346 (1994).

**64.** Mac Lane, S.: *Categories for the Working Mathematician*, GTM vol.5, Springer-Verlag, New York, 1971.

**65.** Mantegna, R.N., Stanley, H.E.: *An Introduction to Econophysics: Correlations and Complexity in Finance*, Cambridge University Press, Cambridge, 2000.

**66.** Melander, M.V., Hussain, F.: Coupling Between a Coherent Structure and Fine-Scale Turbulence, Phys. Rev. E **48**, 4, 2669-2689 (1993).

**67.** Myers, D.: Lindenbaum-Tarski Algebras, in Monk, J.D., Bonnet, R. (eds) *Handbook of Boolean Algebras*, chap. 26, Elsevier Science Publisher B.V., 1989.

**68.** Newcomb, W.A.: Motion of Magnetic Lines of Force, Annals of Physics **3**, 347-385 (1958).

**69.** Noether, A.,E.: *Gesammelte Abhandlungen, (Collected Papers)*, Springer-Verlag, Berlin-New York, 1983.

**70.** Nyikos, P., Reichel, H.C.: On the Structure of Zerodimensional Spaces, Indagationes Math. **37**, 120-136 (1975).

**71.** Parry, W., Pollicott, M.: An Analogue of the Prime Number Theorem for Closed Orbits of Axiom A Flows, Ann. of Math. **118**, 573-591 (1983).

**72.** Peruzzi, A.: The Theory of Descriptions Revisited, Notre Dame J. Formal Logic **30**, 1, 91-104 (1989).

**73.** Reynolds, W.C.: Fundamentals of Turbulence for Turbulence Modeling and Simulation, AGARD-R-**755**, 1.1-1.65 (1987).

**74.** Riccardi, C., Fredriksen, A.: Waves and Coherent Structures in the Turbulent Plasma of a Simple Magnetized Torus, Phys. Plasmas **8**, 1, 199-209 (2001).

**75.** Rieger, L.: *Algebraic Methods of Mathematical Logic*, Academic Press, New York and London, 1967.

**76.** Robert, A.M.: Euclidean Models of p-adic Spaces, in Schikhof, W.H. et al. (eds) *p-adic Functional Analysis*, Marcel Dekker, New York, 1997.

**77.** Robert, A.M.: *A Course in p-adic Analysis*, GTM vol. 198, Springer-Verlag, New York, 2000.

**78.** Ruelle, D.: *Dynamical Zeta Functions for Piecewise Monotone Maps of the Interval*, CRM Monograph Series vol. 4, American Mathematical Society, Rhode Island, 1994.

**79.** Saffman, P.G.: *Vortex Dynamics*, Cambridge Monographs on Mechanics and Applied Mathematics, Cambridge University Press, Cambridge, 1995.

**80.** Sarmant, M.-C., Escassut, A.: Prolongement Analytique A Travers Un T-Filtre, Studia Scientiarum Mathematicarum Hungarica **22**, 407-444 (1987).

**81.** Schikhof, W.H.:*Ultrametric Calculus. An Introduction to p-adic Analysis*, Cambridge stud. adv. math. vol.4, Cambridge University Press, Cambridge, 1984.

**82.** Serre, J.-P.: Groupes de Lie l-Adiques Attachés aux Courbes Elliptiques, in Krasner, M. (ed) *Les Tendances Géométriques en Algèbre et Théorie des Nombres*, Éditions du CNRS, Paris, 1966.

**83.** Serre, J.-P.: *A Course in Arithmetics*, GTM vol. 7, Springer-Verlag, New York, 1996.

**84.** Silverman, J.H.: *The Arithmetic of Elliptic Curves*, GTM vol. 106, Springer-Verlag, New York, 1986.

**85.** Silverman, J.H.: *Advanced Topics in the Arithmetic of Elliptic Curves*, GTM vol 151, Springer-Verlag, New York, 1994.

**86.** Siregar, E., Stribling, W.T., Goldstein, M.L.: On the Dynamics of a Plasma Vortex Street and its Topological Signatures, Phys. Plasmas **1**, 2125-2134 (1994).

**87.** Sirovich, L.: Turbulence and the Dynamics of Coherent Structures, Part I: Coherent Structures, Quart. Appl. Math. **45**, 3, 561-571 (1987).

**88.** Sirovich, L.: Turbulence and the Dynamics of Coherent Structures, Part II: Symmetries and Transformations, Quart. Appl. Math. **45**, 3, 573-582 (1987).

**89.** Smale, S.: Differentiable Dynamical Systems, Bull. Amer. Math. Soc. **73**, 747-817 (1967).

**90.** Sornette, D., Johansen, A., Bouchaud, J.-P.: Stock Market Crashes, Precursors and Replicas, J. Phys. I France **6**, 167-175 (1996).

**91.** Sreenivasan, K.R.: Fluid Turbulence, Rev. Mod. Phys. **71**, 2, S383-S395 (1999).

**92.** Stanley, H.E.: Exotic Statistical Physics: Applications to Biology, Medecine, and Economics, Physica A **285**, 1-17 (2000).

**93.** Stiller, P.F.: Differential Equations Associated With Elliptic Surfaces, J. Math. Soc. Japan **32**, 2, 203-233 (1981).

**94.** Stiller, P.F.: *Automorphic Forms and the Picard Number of an Elliptic Surface*, F. Vieweg & Sohn, 1984.

**95.** Stone, M.H.: The Theory of Representations for Boolean Algebras, Trans. Amer. Math. Soc. **40**, 37-111 (1936).

**96.** Stone, M.H.: Applications of the Theory of Boolean Rings to General Topology, Trans. Amer. Math. Soc. **41**, 375-481(1937).

**97.** Stone, M.H.: Algebraic Characterization of Special Boolean Rings, Fundamenta Math. **XXIX**, 223-303 (1937).

**98.** Sweet, A.M.: A Pragmatic Theory of Locally Standard Grammar, Notre Dame J. Formal Logic **25**, 4, 364-382 (1984).

**99.** Sweet, A.M.: *The Pragmatics and Semiotics of Standard Languages*, Pennsylvania State University Press, University Park and London, 1988.

**100.** Sweet, A.M.: Nominalization in Locally Standard Grammar, in Martín-Vide, C. (ed) *Current Issues in Mathematical Linguistics*, Elsevier Science Publishers B.V., 1994.

**101.** Sweet, A.M.: Local Semantic Closure, Linguistics and Philosophy **22**, 509-528 (1999).

**102.**. Tate, J.T.: Rigid Analytic Spaces, Inventiones math. **12**, 257-289 (1971).

**103.** Tate, J.T.: The Arithmetic of Elliptic Curves, Inventiones math. **23**, 179-206 (1974).

**104.** Taylor,R., Wiles, A.: Ring Theoretic Properties of Certain Hecke Algebras, Ann. of Math. **141**, 553-572 (1995).

**105.** Thiran, E., Verstegen, D., Weyers, J.: p-adic Dynamics, J. Stat. Phys. **54**, 893-913 (1989).

**106.** Wiles, A.: Modular Elliptic Curves and Fermat's Last Theorem, Ann. of Math. **141**, 443-551 (1995).

**107.** Zagier, D.: Introduction to Modular Forms, in Waldschmidt, M. et al. (eds) *From Number Theory to Physics*, Springer-Verlag, Berlin, Heidelberg, 1995.

## 8. Appendix (an informal glossary).

Absolute j-invariant of an elliptic curve - Also called modular invariant, or just j-invariant; it classifies elliptic curves up to isomorphism. Semi-stable elliptic curves with different j-invariants are isomorphic to different analytic tori (see [11], pp. 414-415, [45] and [18]).

Addition law of an elliptic curve - A composition law, defined on the set of points of an elliptic curve with a specified rational point 0, and turning it into an abelian group with identity 0.

Affinoid variety - An affinoid algebra, together with the set of all its maximal ideals (an affinoid algebra is a Banach algebra, together with a map from, at least, one Tate algebra to that same Banach algebra).

Analytic torus - A rigid analytic variety that is obtained by pasting together some affinoid subdomains of the unit disc, with this later being understood as an affinoid variety (see [11], pp. 364-365).

Axiom A - We say that a given diffeomorphism, $f$, of a compact manifold, M, obeys Axiom A when its set of nonwandering points is an hyperbolic one. Under this condition, it follows that the number of isolated periodic points of period n of $f$ satisfies a certain estimate concerning its growth with n (see [89], pp. 764-766 and 777). Now, it happens that this estimate has some interesting models in arithmetic categories, as these models are used as asymptotic equi-distribution axioms counting the number of elements of norm not exceeding a certain value, in each equivalence class of any given arithmetical formation (see [54], p. 264).

Conductor of an elliptic curve - An isogeny invariant of the curve.

Cusp newform of weight two and level N - An element of the set of holomorphic functions (called modular forms) satisfying a particular (weight two) version of the modular transformation property associated with the congruence subgroup of level N. The 'cusp' qualification comes from a special condition on the coefficients of the associated expansion at infinity; the further qualification, from the affix 'new', comes from the relation between these coefficients and the rational eigenvalues of Hecke operators acting on its space.

Dirichlet character modulo m - An arithmetical function that is completely multiplicative and periodic with period m, and that takes the value zero at all numbers not relatively prime to m (see [3], p. 138).

Elliptic curve over Q - A non-singular cubic whose Weierstrass form coefficients are in Q; a curve given by a cubic equation in two variables, with rational coefficients.

Elliptic surface - A surface with a fibration onto a non-singular curve, such that the generic fibre is an irreducible elliptic curve (see the remarks in [85], p. 203).

Euler product (associated with an elliptic curve) - The product of all the local L factors associated with an elliptic curve.

Euler totient function - A function that, to each positive integer, n, associates the number of positive integers not exceeding n and relatively prime to n (see [3], pp. 25-28).

Global L series of an elliptic curve - The same as its Euler product; sometimes written additively, as a Dirichlet series. Also refered to as L function of an elliptic curve.

Global minimal model of an elliptic curve - A Weierstrass equation modeling an elliptic curve, with integer coefficients, and such that, for all primes p, the power of p dividing the discriminant of the curve cannot be decreased by an admissible change of variables (see [51], pp. 290-294, or [84], pp. 224-227).

Infraconnected set - a set, D, that does not admit an empty annulus; given a,x ∈ D, arbitrary, the mapping $\varphi_a(x) = |x-a|$ has an image whose closure in $]0,+\infty[$ is an interval (see [26], chapter 2).

Lindenbaum-Tarski algebra - It is the algebra of the equivalence classes of sentences with respect to a given theory (see [57,67,75]).

Local L factor - A formal power series recording the reduction type of an elliptic curve at the prime p.

Magnetic confinement chamber of the tokamak type - A device built to achieve the confinement of a plasma, using magnetic fields to give it the topology of a torus. One of the most promising machines conceived to achieve controlled nuclear fusion.

Magnetic induction field - The field that gives origin to the forces used to confine the electrically charged particles in a plasma. It is a divergence free field that can be expressed as the curl of a magnetic vector potential depending on the electric currents.

Magnetic flux surface - A toroidal surface that is covered by the field lines of the magnetic induction field.

Magnetic trigger mode - The sudden jump or increase of the growth rate of a fluctuation in the plasma (see [46] ).

Modular parametrization of level N - A non-constant morphism, from the compactification of the quocient of the upper half complex plane by the Hecke congruence subgroup of level N, to an elliptic curve over Q.

Non-archimedean valuation - A valuation such that $|2| \leq 1$ (see [81], p. 18); one associated with a norm satisfying the ultrametric inequality ( $|x + y| \leq \max(|x|,|y|)$ ).

Non-principal Dirichlet character modulo m - A Dirichlet character modulo m that is not identically 1 at all numbers relatively prime to m (see [3], p. 138).

Numerical stability – let g be any map from a compact differential manifold without boundary, M, to itself. Let f : M→M. Let $N_v(f)$ denote the number of isolated periodic points of period ν of f. Then, f is said to be numerically stable if it has a neighborhood such that, for ν ≥ 1 , $N_\nu(f) \leq N_\nu(g)$ (see [6]).

p-adic valuation - A discrete, non-archimedean valuation (see [81], pp. 8-11).

Plasma - An ionized gas whose constituent particles show complex, collective behaviour. The most common state of matter in the Universe.

Plasma collapse - A catastrophic event, consisting of an abrupt change of the global parameters of a plasma, that terminates the discharge (see [8]).

Polyadic algebra - a structure that stands in relation to the first-order functional calculus like a Boolean algebra does in relation to the propositional calculus. It can be obtained endowing an underlying Boolean algebra with two mutually compatible structures, namely, those of a quantifier algebra and of a transformation algebra.

Polyadic dilation - It is an embedding process of a polyadic algebra into another, which can be thought as a process of adjunction of variables to the former, thereby obtaining the later [21,35].

Reduction type of an elliptic curve - It depends on what results, after having reduced, modulo a prime p, the coefficients of the corresponding global minimal model. We say that the reduction is stable at p if the reduced curve is non-singular; we say that the reduction is unstable at p if the reduced curve has a cusp; we say that the reduction is semi-stable at p if the reduced curve has a node (see [84], pp. 179-183).

Rigid analytic variety - A set endorsed with a specified Grothendieck topology and with a sheaf of rings on it, all of whose stalks are local rings. The set, and the topology on it are assumed to satisfy some additional conditions (see [11], p. 357).

Semi-stable elliptic curve over Q - One whose reduction modulo a prime is a curve with a node; one that is isomorphic to an analytic one-dimensional torus (see [11], pp. 407-415); one that is modular (see [104,106]).

Spherically complete field – a field such that, in it, each nested sequence of discs has a nonempty intersection (see [81], section 20).

T-filter – a filter whose characteristic property is that there exist analytic elements strictly vanishing along it (see [25], or [26], chapters 35 to 38).

Topological duality (Stone duality) - It is an equivalence between the categories of Boolean algebras and zero-dimensional, compact Hausdorff spaces.

Transport barrier - A magnetic flux surface that prevents the electrically charged particles in the plasma of escaping from the region of confinement, and of collision with the walls of its container (the tokamak).

Underlying affinoid structure of an analytic torus - The affinoid subdomains from which the analytic torus is obtained by a pasting procedure (see [11], pp. 364-365).

Vorticity - A vector field associated with the rotational motion of an element of volume of a fluid about an instantaneous axis (see [73], pp. 1.5-1.8, or [79], chapter 1). It is a divergence free field.